\documentclass[10pt,twocolumn,letterpaper]{article}

\usepackage{cvpr}              % To produce the CAMERA-READY version
\usepackage[accsupp]{axessibility} % Improves PDF readability for those with visual impairments.

\usepackage{multirow}
\usepackage[dvipsnames]{xcolor}

\definecolor{cvprblue}{rgb}{0.21,0.49,0.74}
\usepackage[pagebackref,breaklinks,colorlinks,citecolor=cvprblue]{hyperref}

\def\paperID{11047} % *** Enter the Paper ID here
\def\confName{CVPR}
\def\confYear{2024}

\title{
Steganographic Passport: An Owner and User Verifiable Credential for Deep Model IP Protection Without Retraining}

\author{Qi Cui\footnotemark[1]\hspace{1cm} Ruohan Meng\footnotemark[1]\hspace{1cm} Chaohui Xu\hspace{1cm} Chip-Hong Chang\footnotemark[2]\\
School of Electrical and Electronic Engineering, Nanyang Technological University\\
{\tt\small qi.cui@ntu.edu.sg, ruohan.meng@ntu.edu.sg, chaohui001@e.ntu.edu.sg, echchang@ntu.edu.sg}
}

\begin{document}
\maketitle
\begin{abstract}
Ensuring the legal usage of deep models is crucial to promoting trustable, accountable, and responsible artificial intelligence innovation. Current passport-based methods that obfuscate model functionality for license-to-use and ownership verifications suffer from capacity and quality constraints, as they require retraining the owner model for new users. They are also vulnerable to advanced Expanded Residual Block ambiguity attacks.
We propose Steganographic Passport, which uses an invertible steganographic network to decouple license-to-use from ownership verification by hiding the user’s identity images into the owner-side passport and recovering them from their respective user-side passports. An irreversible and collision-resistant hash function is used to avoid exposing the owner-side passport from the derived user-side passports and increase the uniqueness of the model signature.
To safeguard both the passport and model’s weights against advanced ambiguity attacks, an activation-level obfuscation is proposed for the verification branch of the owner’s model.
By jointly training the verification and deployment branches, their weights become tightly coupled.  
The proposed method supports agile licensing of deep models by providing a strong ownership proof and license accountability without requiring a separate model retraining for the admission of every new user. Experiment results show that our Steganographic Passport outperforms other passport-based deep model protection methods in robustness against various known attacks. 
\end{abstract}
    
\vspace{-0.6cm}
\section{Introduction}\label{sec:intro}
\vspace{-0.2cm}
\footnotetext{This research is supported by the National Research Foundation, Singapore, and Cyber Security Agency of Singapore under its National Cybersecurity Research \& Development Programme (Development of Secured Components \& Systems in Emerging Technologies through Hardware \& Software Evaluation \textless NRF-NCR25-DeSNTU-0001\textgreater). Any opinions, findings and conclusions or recommendations expressed in this material are those of the author(s) and do not reflect the view of National Research Foundation, Singapore and Cyber Security Agency of Singapore.}
\footnotetext[2]{Corresponding author; \quad *{Both authors contributed equally.}}

\begin{figure}[t!]
\vspace{-0.6cm}
\centering  \includegraphics[width=0.99\linewidth]{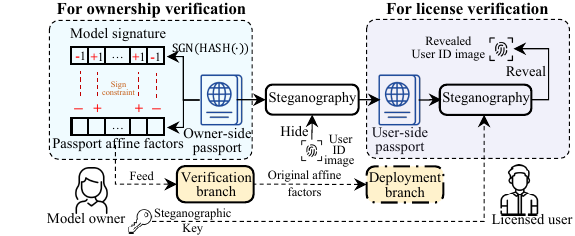}
\vspace{-0.4cm}
     \caption{The conceptual overview. The owner holds the owner-side passport and the steganographic key, which are used to verify the licenses and reveal the user IDs, respectively. The verification branch of the deep model is trained with the constraint of the signature. Upon licensing, the user receives the user-side passport and the deployment branch of the deep model. }
   \label{fig:onecol}
   \vspace{-0.6cm}
\end{figure}
To foster Artificial Intelligence (AI) diffusion and application, AI as a Service (AIaaS) has emerged as a lucrative business model that offers across-the-broad access to high-performance machine learning models with subscription- or transaction-based payment schemes. The customers can submit inputs to the remote server and receive the inference or classification results of machine learning processes through application program interfaces (APIs). As complex models of high capabilities in AIaaS are offered as ``turnkey'' solutions to potentially anyone, they have huge potential to be misused and abused. Ownership and user license verifications are the first step towards accountability of responsible use of AI services~\cite{liu2024threats}. 

Ownership verification has long been a primary tool to protect deployed deep models against intellectual property (IP) infringement and unauthorized model replication~\cite{javadi2021monitoring}. Numerous approaches have been proposed, such as watermarking~\cite{uchida2017embedding, guan2020reversible}, fingerprinting\cite{he2019sensitive, quan2023fingerprinting, peng2022fingerprinting}, and backdoor embedding~\cite{shafieinejad2021robustness, hua2023unambiguous, jia2021entangled}. These approaches typically aim to enable the trained deep model to exhibit distinguishable characteristics under specific conditions without sacrificing its utility. To resist ambiguity attacks, passport-based ownership verification scheme~\cite{fan2019rethinking, fan2021deepipr, zhang2020passport, liu2023trapdoor} embeds digital passports during model training to obfuscate the primary functionality until the obfuscated parameters are restored by a genuine passport. The passport scheme provides access control~\cite{fan2019rethinking, fan2021deepipr} at the expense of some accuracy degradation. Each passport is tied to a user model, \textit{retraining} is required to derive a new user model from the owner model for every new passport issued. The recent passport-aware normalization~\cite{zhang2020passport} avoids making structural changes to the target model by training it jointly with a passport-aware branch. The secret passport and passport-aware branch after training are kept by the model owner for future ownership verification and only the original passport-free target model is released to the end users. By forgoing the usage control, this method allows one deployed model to be used by various users but it also loses the accountability of inappropriate and unlawful use of the deployed model by different licensees. Besides, it has been demonstrated that the passports of existing methods can still be counterfeited by the more advanced ambiguity attack~\cite{chen2023effective}.

In usage authorization, one major limitation of existing passport-based methods is the need to retrain the network to incorporate different parameter modulations for every new passport issued, which may cause uneven service quality for different end users. 
Besides, retraining to accommodate new user models is inflexible, non-scalable and not cost-effective for complex models. This pitfall motivates us to explore anti-forgery passports that \textit{allow fidelity verification of deployed model to hold authorized users accountable for misuse of their licensed models without having to retrain the owner model when admitting new users}. To this end, we propose a novel Steganographic Passport to decouple user license verification from ownership verification, as illustrated in~\cref{fig:onecol}.
For \textit{ownership verification}, the uniqueness of the model weights is maintained by aligning them with the hashed output of the owner-side passport, which allows the model owner to prove the ownership and authenticity of the distributed model. For \textit{license verification}, any licensed user’s hidden identity (ID) image can be extracted using the user-side passport. Both owner-side and user-side passports will pass the fidelity assessment as their visual similarity is preserved by using steganography to hide the ID image. 
To increase the sensitivity of the deployed model against malicious tampering of passports and weights, we propose an activation-level obfuscation for the verification branch of the owner model. The weights of the deployment and verification branches are tightly coupled by training them jointly with a balance loss function. Our contributions can be summarized as follows.
\begin{itemize} 
\item{A novel Steganographic Passport is designed based on key-based invertible steganographic network to allow both original model ownership and right-of-use of its deployed model by multiple licensees to be verified without requiring retraining.}
\item{An activation-level obfuscation architecture is designed to thwart forgery by tampering with the user-side passports or model weights. }
\item{Extensive experiments are conducted to demonstrate the state-of-the-art security performance of our steganographic passports—the creditability of its owner-side and user-side passport verifications and its robustness against various attacks, including ownership ambiguity attacks, license ambiguity attacks, and removal attacks.}
\end{itemize}

\vspace{-0.2cm}
\section{Background}\label{sec:bg}
\vspace{-0.2cm}
The need to safeguard the deep model against misappropriation and IP theft arises from the enormous cost of designing and training a high-performance model from scratch. Given redundancy inherent in the numerous parameters of deep models, it is possible to embed a watermark into a model for ownership identification without significantly impairing its performance. The model can be watermarked either explicitly or implicitly. Explicit watermarking embeds the watermark by modifying the weights~\cite{uchida2017embedding, feng2020watermarking, he2019sensitive, peng2022fingerprinting} or feature maps~\cite{guan2020reversible} through a weight regularizer without compromising the original model performance. Implicit watermarking (a.k.a prediction watermarking) deliberately trains a backdoor into the protected model to classify inputs with specific triggers to the predefined labels~\cite{shafieinejad2021robustness, hua2023unambiguous, wang2022buyer, jia2021entangled, ong2021protecting}. However, these techniques are found to be vulnerable to ambiguity attacks as the embedded triggers can be replaced by attackers to forge the ownership with a substitute watermark. To mitigate this risk, model passports are introduced~\cite{fan2019rethinking, fan2021deepipr}, where the affine scale and bias factors of normalization functions are substituted with those derived from the passport images $\{p_\gamma, p_\beta\}$.

Let $x^{l}$ be the output of the $l$-th convolutional layer, and $\mu_{x^{l}}$ and $\sigma_{x^{l}}$ be its mean and standard deviation, respectively. Then, the normalization function at the $l$-layer with affine factors $\gamma^l$ and $\beta^l$ is given by:
\vspace{-0.2cm}
\begin{equation}\label{eq:norm_general}
\begin{aligned} 
    \widehat{x^{l}} &= \gamma^l \cdot \tilde{x^l} + \beta^l,\\  
    \text{where} \quad \tilde{x^l} &= \frac{1}{\sigma_{x^{l}}}(x^{l}-\mu_{x^{l}}).
\end{aligned}
\vspace{-0.3cm}
\end{equation}

By substituting $\gamma^l$ and $\beta^l$ by the passport affine factors $\gamma^l_{p_\gamma}$ and $\beta^l_{p_\beta}$, the following passport normalization~\cite{fan2019rethinking} is obtained:
\begin{equation}\label{eq:passport_general}
\begin{aligned} 
    \widehat{x^{l}_p} = \gamma^l_{p_\gamma} \cdot \tilde{x^l} + \beta^l_{p_\beta}.
\end{aligned}
\end{equation}

\begin{figure}[t]
  \centering
  \vspace{-0.8cm}
  \begin{subfigure}{0.236\textwidth} 
\includegraphics[width=\textwidth]{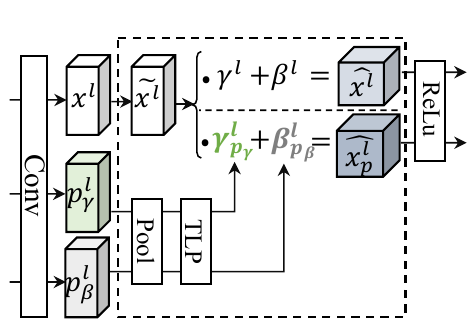}
        \caption{Existing passport architecture}
        \label{fig:strc-a}
    \end{subfigure}
    \begin{subfigure}{0.236\textwidth}
\includegraphics[width=\textwidth]{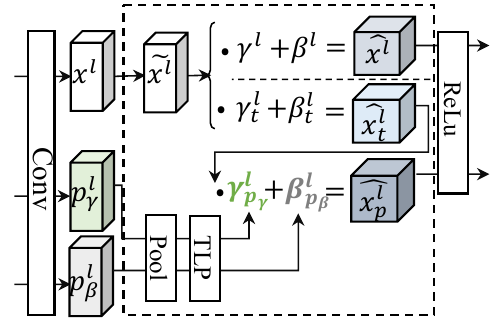}
       \caption{Proposed passport architecture}
       \label{fig:strc-b}
    \end{subfigure}
  \vspace{-0.7cm}
   \caption{The existing and proposed passport architectures. The dual branches are enclosed in the dashed line box. }
   \label{fig:strc}
   \vspace{-0.6cm}
\end{figure}

The sign sequence of $\gamma^l_{p_\gamma}$ in~\cref{eq:passport_general} is constrained to match a self-defined binary signature sequence. As large magnitudes of $\gamma^l_{p_\gamma}$ can be penalized,  the gradients are kept small to ensure that the signs of $\gamma^l_{p_\gamma}$ are lazy-to-flip so as to prevent the signature from being replaced by retraining the model. To minimize the influence of passport normalization on inference performance, Zhang \textit{et al.}~\cite{zhang2020passport} proposed a dual-branch passport structure. Its normalization function is expressed in~\cref{eq:passport_aware} and illustrated in~\cref{fig:strc-a}.
\vspace{-0.1cm}
\begin{equation}\label{eq:passport_aware}
\vspace{-0.1cm}
\begin{aligned} 
    \left\{\begin{array}{l}
    \widehat{x^{l}}=\gamma^l \cdot \tilde{x^l}+\beta^l,\\
    \widehat{x^{l}_p}=\gamma^l_{p_\gamma} \cdot \tilde{x^l}+\beta^l_{p_\beta},
\end{array}\right.
\end{aligned}
\vspace{-0.2cm}
\end{equation}
where the passport-free output $\widehat{x^{l}}$ and the passport-aware output $\widehat{x^{l}_p}$ are used for the deployment and ownership verification, respectively. Besides, a two-layer perception (TLP) structure is also inserted before the affine factors to mitigate the strict limitations in managing the functionality of normalization layers.

To protect against ambiguity attacks with oracle passports due to passport leakage, Liu \textit{et al.}~\cite{liu2023trapdoor} used trapdoor normalization to establish a direct correlation between the signature and the passport through a hash function. The collision-intractable hash function makes it virtually impossible for the attackers to reverse engineer a counterfeit passport, even if they can compromise the signature. Recently, Chen \textit{et al.}~\cite{chen2023effective} proposed an advanced ambiguity attack called Expanded Residual Block (ERB) on passport-based methods. This attack exploits the extensive solution space of TLP to force-flip the signs of the scale factors according to their forged passport without compromising the model performance.
In summary, existing model passport methods aim only to resist the diverse ambiguity attacks by obfuscating the normalization functionalities with immutable passports. They do not allow admission of new users and their right-of-use verification upon model deployment. This is an important measure to safeguard the interest of AIaaS providers against revenue extorts by unauthenticated usage. 

\vspace{-0.3cm}
\section{Proposed method}
\vspace{-0.1cm}
\subsection{Problem statement }\label{sec:problem}
\vspace{-0.2cm}
We consider a practical licensing scenario of a deep model. Besides being able to detect and verify the ownership of a deployed model, the model owner should also be able to verify usage permission. If the passport images $\{p_\gamma, p_\beta\}$ are used as a \textit{unique} proof to verify the ownership of the deployed model, they cannot be used to verify the usage permission granted to licensed users of that model. Although new backdoor triggers or watermarks can be added to verify each licensed user, the model needs to be retrained to learn the new features. This is very restrictive and impractical as it requires the model owner to fix the number of users and their identities upfront. Therefore, our aim is to enable the passport-based scheme to grant verifiable usage permission of a deployed deep model to ad hoc registered users \textit{without requiring retraining} and \textit{without undermining the validity and robustness of the model’s ownership proof}.

\vspace{-0.1cm}
\subsection{Steganographic passport framework}
\vspace{-0.2cm}
Recall that in existing passport approaches~\cite{fan2019rethinking, fan2021deepipr, zhang2020passport, liu2023trapdoor}, a forged passport image can also be used to pass the fidelity assessment if it bears a high visual similarity to the genuine passport image. Inspired by advanced steganography techniques, we apply the art of imperceptible secret concealment to the passport-based scheme. The proposed Steganographic Passport follows the dual-branch setting, as illustrated in~\cref{fig:strc-b}.

We define an \textit{owner-side passport} $p_o=\{p_{o_\gamma}, p_{o_\beta}\}$ as the genuine passport images selected and held only by the model owner for generating the affine scale and bias factors accordingly. By feeding $p_o$ into the model, the corresponding owner-side passport feature maps are sequentially generated across the $n$ passport layers. These feature maps are then transformed into the affine factors for obfuscating the verification branch, which will be described in~\cref{sec:act_obfuscat}. We also introduce a new \textit{user-side passport}, $p_u=\{p_{u_\gamma}, p_{u_\beta}\}$, which is obtained by hiding the ID image $\mathrm{I}_u$ provided by a registered user into $p_o$. The user-side passport $p_u$ fills the gap of existing passport-based limitation in meeting the goal of~\cref{sec:problem}. The model owner can license the deployment branch of the model to multiple registered users by merely hiding each user’s ID image into $p_o$.

\textbf{Anti-forgery owner-side passport.}
To avoid user-side passports from exposing the visual content of the owner-side passports, the owner-side passport is irreversibly transformed to a unique model signature to ensure that the ownership cannot be forged by knowing any of the user-side passports. SHA-512~\cite{sanadhya2008new} is used as the one-way function for this mapping.

Before training the target model, the model owner hashes the owner-side $p_{o_\gamma}$ to a binary signature $\xi^l$ of passport layer $l$, as follows:
\vspace{-0.2cm}
\begin{equation}\label{eq:hash}
    \xi^l=\{\operatorname{SGN}(x) \mid x \in \operatorname{HASH}(p_{o_\gamma})\},
    \vspace{-0.2cm}
\end{equation}
where $\operatorname{SGN}$ maps each bit of a binary input sequence from $\{0, 1\}$ to $\{-1, 1\}$. If the desired length of $\xi^l$ is shorter than 512 bits, the excess bits of the hash output are truncated from the beginning; Otherwise, we cycle the hash function's output until the required sequence length is obtained. 

For a model with $n$ passport layers, its overall model signature is represented as $\{\xi^1, \xi^2, ..., \xi^n\}$. The strict avalanche criterion of the hash function ensures that even minor changes to the passport image will result in significantly different affine factors, which prevents force-flipping from forging the passport with high model performance fidelity. Together with the irreversibility property of SHA-512, the genuine model signature $\xi$ can only be generated with the owner-side passport to thwart ownership impersonation fraud. The signature $\left\{\xi^1, \xi^2, \ldots, \xi^n\right\}$ will be used to constrain the model’s learning in the subsequent training phase to transform them into unique model's weights. A detailed exposition of this process is provided in~\cref{sec:act_obfuscat}.

\textbf{License verification via key-based deep steganography.}
To support an agile licensing scheme that is capable of admitting and verifying any new user after the target model has finalized its training phase, we need to decouple the license verification from the ownership verification. To this end, we hide the user’s ID into the owner-side passport by integrating the key-based scheme of Invertible Steganographic Network (ISN)~\cite{lu2021large} within the Deep Image Hiding Network (HiNet) framework~\cite{jing2021hinet}. The key-based ISN is divided into two processes: forward hiding pass $\mathcal{H}_{s}$ and reverse revealing pass $\mathcal{H}^{-1}_{s}$. The forward hiding pass is used to obtain the user-side passport $p_u$ from the owner-side passport $p_o$ as follows:
\vspace{-0.2cm}
\begin{equation}\label{eq:hiding}
    p_u = \mathcal{H}_{s}(p_o, \mathrm{I}_u),
    \vspace{-0.2cm}
\end{equation}
where $\mathrm{I}_u$ denotes the user’s ID image to be hidden into the owner-side passport $p_o$.

The reverse revealing pass is utilized for license verification. Using the private steganographic key $\mathbf{k}_s$, the model owner can reveal the hidden user’s ID image from the user-side passport $p_u$:
\vspace{-0.2cm}
\begin{equation}\label{eq:revealing}
    \mathrm{I}^{\prime}_u = \mathcal{H}^{-1}_{s}(p_u, \mathbf{k}_s),
    \vspace{-0.2cm}
\end{equation}
where $\mathrm{I}^{\prime}_u$ denotes the revealed ID image of the licensed user. The detailed architecture and algorithms of the hiding and revealing passes are given in the Appendix.

\subsection{Activation-level obfuscation}\label{sec:act_obfuscat}
\vspace{-0.2cm}
To counteract sophisticated ambiguity attacks, an activation-level obfuscation is proposed to safeguard the verification branch against alterations of both the passport and the model’s weights. 

\textbf{Immutable passport.}
As illustrated in~\cref{fig:strc-b}, the passport-affine factors of layer $l$ can be calculated from the passport feature maps ${p_o^l}$ by:
\vspace{-0.2cm}
\begin{equation}\label{eq:passport_derived}
\begin{aligned} 
    \gamma_{p_{o_\gamma}}^l = \Phi\left(\Pi\left( p_{o_\gamma}^l \right) \right), \mbox{and } 
    \beta_{p_{o_\beta}}^l = \Phi\left(\Pi\left( p_{o_\beta}^l \right) \right),
\end{aligned}
\vspace{-0.2cm}
\end{equation}
where $\Phi$ and $\Pi$ denote the TLP and the global average pooling, respectively.

Inspired by the controllable rectifiers of the Dynamic ReLU~\cite{chen2020dynamic}, we make $\gamma^l_{p_{o_\gamma}}$ and $\beta^l_{p_{o_\beta}}$ the dependent variables of the activation function. This can be incorporated into a dual-branch architecture as follows:
\vspace{-0.2cm}
\begin{equation}\label{eq:obfuscation}
\begin{aligned} 
    \left\{\begin{array}{l}
    \widehat{x^{l}}=\gamma^l \cdot \widetilde{x^{l}}+\beta^l,\\
\widehat{x_{p_o}^{l}}=\gamma^l_{p_{o_{\gamma}}} \cdot (\gamma_t^l \cdot \widetilde{x^{l}}+\beta_t^l)+\beta^l_{p_{o_{\beta}}},
\end{array}\right.
\end{aligned}
\vspace{-0.2cm}
\end{equation}
where $\gamma^l$ and $\beta^l$ denote the original affine factors used in the deployment branch, and $\gamma_t^l$ and $\beta_t^l$ denote the additional temporal affine factors used in the verification branch.  

To simplify the expressions, let $\widehat{x_t^{l}} = \gamma_t^l \cdot \widetilde{x^{l}}+\beta_t^l$. By subjecting the outputs of~\cref{eq:obfuscation} to an activation function (\textit{e.g.} ReLU) before feeding them to the next layer, the final outputs of layer $l$ are given by:
\vspace{-0.2cm}
\begin{equation}\label{eq:obfuscation_relu}
\begin{aligned} 
    \left\{\begin{array}{l}
    x_{\varphi}^{l}=\max(\widehat{x^{l}}, 0),\\
    x^{l}_{\varphi_{p_o}}=\max(\gamma^l_{p_{o_{\gamma}}} \cdot \widehat{x_t^{l}}+\beta^l_{p_{o_{\beta}}}, 0).
\end{array}\right.
\end{aligned}
\vspace{-0.2cm}
\end{equation}

The activation function used for the deployment branch (top expression) in~\cref{eq:obfuscation_relu} is a regular ReLU, whereas the activation function used for the verification branch (bottom expression) can be viewed as a special case of Dynamic ReLU. This way, the changes in the affine factors due to any alterations of the owner-side passport $p_o$ will lead to apparent changes in the passport-aware output, $x_{\varphi_{p_o}}^{l}$, but not in the passport-free output, $ x_{\varphi}^{l}$.

\textbf{Immutable deployment branch.}
Under the dual-branch setting, the deployment and verification branches share all the model’s weights but the affine factors. The deployment branch has to be consistently updated according to the updated passport-derived affine factors in the verification branch. If the deployment branch cannot keep up with the updates, the shared weights will overfit the verification branch, causing gradient explosions. To keep the updates of the two branches in tandem, the following balance loss function is proposed to train the model by explicitly aligning the affine factors in~\cref{eq:obfuscation}:
\vspace{-0.2cm}
\begin{equation}\label{eq:balloss}
\mathcal{L}_{b} \!= \!\sum_{l=1}^n \! \left(\! \mathcal{L}_{\emph{y}}(\gamma^l, \gamma_t^l\! \cdot \!\gamma_{p_{o_{\gamma}}}^l) \!+ \! \mathcal{L}_{\emph{y}}(\beta^l, \beta_t^l\! \cdot \! \gamma_{p_{o_{\gamma}}}^l \!\!+ \! \beta^l_{p_{o_{\beta}}} )\right)\!,
\vspace{-0.2cm}
\end{equation}
where $\ell_1$ norm is used for $\mathcal{L}_{\emph{y}}$, and $n$ is the total number of passport layers. 

This loss function aims to strike an equilibrium performance between the two branches. After the model is trained, the verification branch is removed from the model before it is licensed to any user. As the model trained with the balance loss function will have negligible accuracy difference between the two branches, any modification on the deployment branch would amplify the mean absolute difference. By comparing the accuracy difference between the two branches, the integrity of the model can be validated.

\textbf{Immutable verification branch.}
During the target model’s training phase, the owner uses the model signature $\left\{\xi^1, \xi^2, \ldots, \xi^n\right\}$ to constrain the update of model's weights with the following signature loss function:
\vspace{-0.3cm}
\begin{equation}\label{eq:signloss}
\mathcal{L}_s = \sum_{l=1}^n \max \left(\epsilon-\xi^l \cdot {\Pi}\left( p_{o_\gamma}^l \right) \cdot \gamma_{p_{o_\gamma}}^l, \, 0\right),
\vspace{-0.3cm}
\end{equation}
where $\xi^l$ and $p_{o_\gamma}^l$ denote the $l$-th layer's signature and passport feature map, respectively, and $\epsilon$ is a small constant. 

$\mathcal{L}_s$ encourages the signs of the product ${\Pi}\left( p_{o_\gamma}^l \right) \cdot \gamma_{p_{o_\gamma}}^l$ to align with that of $\xi^l$, while keeping the magnitude of this product low. Since $p_{o_\gamma}^l$ and $\gamma_{p_{o_\gamma}}^l$ are the outputs of the convolution and TLP, respectively, the weights of the convolution layer and TLP become tightly coupled. The uniqueness of $\xi^l$ in~\cref{eq:signloss} contributes to the distinctiveness of the weights. Any adjustment to the TLP will directly affect the convolution output. This property helps to prevent advanced ambiguity attacks, especially ERB, which maliciously alter the weights of the TLP to force-flip the signs of the product to align it with the forged signature.

\subsection{Training objectives}
\vspace{-0.2cm}
The one-time training pipeline consists of training the key-based ISN and the target model.

\textbf{Training the key-based ISN.}
The key-based ISN is trained once independently of the target model to enhance the hiding invisibility and the revealing quality by the following loss function:
\vspace{-0.2cm}
\begin{equation}\label{eq:inn}
\mathcal{L}_{\emph{ISN}} = \mathcal{L}_{h}(p_o, p_u) + \mathcal{L}_{r}(\mathrm{I}_u, \mathrm{I}^{\prime}_u).
\vspace{-0.2cm}
\end{equation}

Both $\mathcal{L}_{h}$ and $\mathcal{L}_{r}$ utilize the $\ell_2$ norm for distance calculation. $p_o$ and $\mathrm{I}_u$ are randomly sampled in the training phase.

\textbf{Training the target model.}
Given a target model, we denote the models with only the weights of the deployment branch as $\mathcal{M}$, and those with only the weights of the ${p_o}$-based verification branch as $\mathcal{M}_{p_o}$. Three loss functions are used to train the dual branches parallelly. The loss function for model performance is defined as:
\vspace{-0.2cm}
\begin{equation}\label{eq:train_per}
\!\!\! \mathcal{L}_{f} \!= \! \frac{1}{N_{tr}} \!\sum_{i=1}^{N_{tr}} \left(\mathcal{L}_{c}\left(\mathcal{M}\left(x_i\right), y_i\right) \!+\! \mathcal{L}_{c}\left(\mathcal{M}_{p_o}\left(x_i\right), y_i\right) \right)\!,
\vspace{-0.2cm}
\end{equation}
% \vspace{-0.2cm}
% \begin{equation}\label{eq:train_per}
% \mathcal{L}_{f} = \mathcal{L}_{c}(\mathcal{M}(x), y) + \mathcal{L}_{c}(\mathcal{M}_{p_o}(x), y),
% \vspace{-0.2cm}
% \end{equation}
where $x_i$ and $y_i$ are respectively the $i$-th sample and its label from $N_{tr}$ training samples, and $\mathcal{L}_{c}$ is the cross-entropy loss.

Together with the signature loss $\mathcal{L}_{s}$ and the balance loss $\mathcal{L}_{b}$, the total loss is defined as:
\vspace{-0.2cm}
\begin{equation}\label{eq:totalloss}
\mathcal{L}_{t} = \mathcal{L}_{f} + \omega_{s}  \mathcal{L}_{s} + \omega_{b} \mathcal{L}_{b},
\vspace{-0.2cm}
\end{equation}
where the signature and balance losses are weighted by $\omega_{s}$ and $\omega_{b}$, respectively.

\subsection{Inclusive verification chain}\label{sec:veri}
\vspace{-0.2cm}
To confirm the ownership and legal usability of a deployed model $\mathbf{M}$, we propose the following verification chain involving the model owner $\mathbf{O}$, the licensed user $\mathbf{U}$, and the deployed model $\mathbf{M}$.

\textbf{Ownership verification.}\label{sec:ownership_veri}
The ownership verification comprises three tests: $\mathcal{V}^\mathcal{I}_{O}$, $\mathcal{V}^f_{O}$ and $\mathcal{V}^s_{O}$ for the verification of the model integrity, performance fidelity and signature accuracy, respectively. By replacing the passport layers' affine factors with the ones derived from the owner-side passport $p_o$, the deployed branch $\mathbf{M}$ becomes a verifiable branch $\mathbf{M}_{p_o}$. 

Let $\mathcal{F}(\cdot,\cdot)$ denote the model performance. $\mathcal{V}^\mathcal{I}_{O}$ tests if $\mathbf{M}$ has been modified based on the absolute performance difference (AD):
\vspace{-0.2cm}
\begin{equation}\label{eq:V_Own_i}
\begin{aligned}
\!\mathcal{V}_{O}^\mathcal{I}  \!\!\! \iff \!\!  &\frac{1}{N_t} \!\left| \sum_{j=1}^{N_t} \left( \mathcal{F} \left( \mathbf{M}_{p_o}, {x}_j^t \right) \!-\! \mathcal{F}\left(\mathbf{M}, {x}_j^t\right) \right) \right| \!\!<\! \mathbf{\tau}_{d}, \!
\end{aligned}
\vspace{-0.2cm}
\end{equation}
where ${x}_j^t$ is the $j$-th sample from total ${N_t}$ test samples, and $\mathbf{\tau}_{d}$ is a small predefined passing threshold. 

$\mathcal{V}^f_{O}$ verifies the performance fidelity by:
\vspace{-0.3cm}
\begin{equation}\label{eq:V_Own_f}
\begin{aligned}
\mathcal{V}^f_{O} \iff \frac{1}{N_t} \sum_{j=1}^{N_t} \mathcal{F} \left( \mathbf{M}_{p_o}, {x}_j^t \right) > \mathbf{\tau}_{f},
\end{aligned}
\vspace{-0.3cm}
\end{equation}
where $\mathbf{\tau}_{f}$ is the predefined minimum accuracy to pass the performance fidelity verification. 

To validate the signature accuracy, the owner uses~\cref{eq:hash} to transform $p_{o_\gamma}$ into the model signature $\xi = \left\{\xi^1, \xi^2, \ldots, \xi^n\right\}$. By comparing each element of $\xi$ with those of $\xi^*$ extracted from $\mathbf{M}_{p_o}$, $\mathcal{V}^s_{O}$ checks the sign agreement (SA) by:
\vspace{-0.4cm}
\begin{equation}\label{eq:V_own_s}
    \mathcal{V}^s_{O} \iff \frac{1}{N_{\xi}} \sum_{b=1}^{N_{\xi}} (\xi_b \land \xi^*_b) \ge \mathbf{\tau}_{\xi},
    \vspace{-0.2cm}
\end{equation}
where ${N_{\xi}}$ is the total length of $\xi$. $\mathbf{\tau}_{\xi}$ is a predefined  error tolerance \textit{only} for the post-distribution verification. Otherwise, the SA of an untainted pre-deployment model has to be 100\% (\textit{i.e.}, $\mathbf{\tau}_{\xi} = 1$) to satisfy $\mathcal{V}^s_{O}$.

\textit{Remark~1:} \textit{Before distributing the deployed model $\mathbf{M}$, the owner $\mathbf{O}$ is required to disclose the evidence of compliance with $\mathcal{V}^\mathcal{I}_{O}$, $\mathcal{V}^f_{O}$, and $\mathcal{V}^s_{O}$ to establish the credibility of the ownership claim of $\mathbf{M}$.}

\textbf{License verification.}\label{sec:license_veri} 
Upon verifying that the model $\mathbf{M}$ deployed by the licensed user $\mathbf{U}$ complies with the criteria set by $\mathcal{V}^\mathcal{I}_{O}$, $\mathcal{V}^f_{O}$, and $\mathcal{V}^s_{O}$, a provable association between the licensed user $\mathbf{U}$ and the model owner $\mathbf{O}$ can be established by two license verification tests: $\mathcal{V}_{L}^p$ for passport image similarity, and $\mathcal{V}_{L}^{\mathrm{I}}$ for ID image similarity. 

Upon receiving $p_u$ submitted by the user $\mathbf{U}$, the owner can compare the PSNR between the user and owner passport images by:
\vspace{-0.2cm}
\begin{equation}\label{eq:V_Lic_p}
    \mathcal{V}_{L}^p \iff \operatorname{PSNR}(p_o, p_u) > \mathbf{\tau}_{p},
    \vspace{-0.2cm}
\end{equation}
where $\mathbf{\tau}_{p}$ is the predefined visual similarity threshold for successful verification. 

The reverse revealing pass is utilized for license verification. Using the private steganographic key $\mathbf{k}_s$, the model owner can reveal the hidden user’s ID image $\mathrm{I}^{\prime}_u$ from the user-side passport $p_u$ by~\cref{eq:revealing}. $\mathcal{V}_{L}^{\mathrm{I}}$ tests the visual similarity between $\mathrm{I}_u$ and the revealed ID image, $\mathrm{I}^{\prime}_u$ of the licensed user by comparing the PSNR with predefined threshold $\mathbf{\tau}_{r}$:
\vspace{-0.2cm}
\begin{equation}\label{eq:verfication_lic}
  \mathcal{V}_{L}^{\mathrm{I}}  \iff \operatorname{PSNR}(\mathrm{I}_u, \mathrm{I}^{\prime}_u) > \mathbf{\tau}_{r}.
  \vspace{-0.2cm}
\end{equation}

\textit{Remark~2:} \textit{Once $\mathcal{V}_{L}^p$ and $\mathcal{V}_{L}^{\mathrm{I}}$ are satisfied, on the condition that the model $\mathbf{M}$ passes the three ownership verification tests of the owner $\mathbf{O}$, the user $\mathbf{U}$ is proven to be a legal licensee of $\mathbf{M}$.}

\vspace{-0.2cm}
\section{Experiment}
\begin{table*}[t!]
\centering
\vspace{-0.8cm}
\caption{{Inference performance (in \%) for deployment/verification. $\uparrow$ / $\downarrow$ denotes higher / lower value for better performance. } }\label{tab:inference}
\vspace{-0.4cm}
\resizebox{\linewidth}{!}{
\begin{tabular}{c  c  c | c  c | c  c | c  c| c c}
\toprule
 & \multicolumn{2}{c|}{CIFAR-10} & \multicolumn{2}{c|}{CIFAR-100} & \multicolumn{2}{c|}{Caltech-101} & \multicolumn{2}{c|}{Caltech-256} &   &  \\
 \hline
 {ResNet-18} & BN$\uparrow$  & GN$\uparrow$ & BN$\uparrow$ & GN$\uparrow$ & BN$\uparrow$ & GN$\uparrow$ & BN$\uparrow$ & GN$\uparrow$ & Mean$\uparrow$ & AD$\downarrow$\\
\hline
Clean  & 94.69 & 93.85 & 76.42 & 73.57 & 74.58 & 70.73 & 55.43 & 49.52 & 73.60& -- \\

DeepIPR & 93.04  /  94.05 & 93.81 / 93.69 & 60.76 / 62.88  & 72.19 / 71.15 & 73.84 / 73.22 & 65.08 / 63.33 & 41.53 / 44.12 & 43.81 / 42.67 & 68.01 / 68.14 & 1.30 \\

PAN & 94.70 / 94.61 & 93.63 / 93.67 & 75.97 / 74.70 & 72.19 / 71.49 & 73.79 / 72.94 & 68.36 / 65.37 & 55.37 / 54.33 & 44.73 / 43.51 &  72.34 / 71.33 & 1.03\\

TdN & 94.65 / 94.68  & 93.57 / 93.57  & 76.00 / 75.61  & 71.25 / 71.42 & 74.12 / 73.05 & 67.63 / 67.51  & 54.90 / 54.28 & 44.90 / 44.27 & 72.13 / 71.80 & 0.38\\

Ours & \textbf{94.89} / \textbf{94.89}  & 93.79 / \textbf{93.80}  & \textbf{76.10} / \textbf{76.10}  & 70.84 / 70.83  & 73.56 / \textbf{73.56}  & 67.01 / 67.01 & 53.65 / 53.65 & 44.01 / 44.01 & 71.73 / 71.73 & \textbf{0.00} \\

\hline
\hline

{AlexNet} & BN$\uparrow$  & GN$\uparrow$ & BN$\uparrow$ & GN$\uparrow$ & BN$\uparrow$ & GN$\uparrow$ & BN$\uparrow$ & GN$\uparrow$ & Mean$\uparrow$ & AD$\downarrow$\\
\hline
Clean  & 91.20 & 90.27 & 68.40 & 65.72 & 71.58 & 68.70 & 44.24 & 41.18 & 67.66& -- \\

DeepIPR  & 81.80 / 90.76 & 89.90 / 90.20  & 45.71 / 51.77 & 64.89 / 64.16 & 67.91 / 65.95 & 67.80 / 67.06 & 31.39 / 36.25 & 40.69 / 39.73  & 60.14 / 63.24 & 3.07 \\

PAN  & 91.51 / 91.29 & 90.22 / 90.39  & 68.46 / 66.44 & 50.60 / 56.71 & 71.03 / 70.90 & 66.95 / 66.71 & 44.93 / 41.13 & 40.03 / 38.84  & 66.47 / 65.31 & 1.32 \\

TdN  & 91.28 / 91.02 & 90.08 / 89.80  & 68.71 / 68.07 & 64.54 / 63.77 & 70.45 / 69.44 & 68.31 / 66.38 & 45.36 / 44.82 & 42.39 / 37.88  & 67.64 / 66.40 & 1.24 \\

Ours  & 91.39 / \textbf{91.38} & 89.77 / 89.77  & 67.86 / 67.85 & 63.36 / 63.34 & \textbf{71.98} / \textbf{71.98} & \textbf{68.42} / \textbf{68.42} & \textbf{46.10} / \textbf{46.07} & 41.10 / 41.12  & 67.50 / \textbf{67.49} & \textbf{0.01} \\

\bottomrule
\end{tabular}
}
\vspace{-0.2cm}
\end{table*}
\subsection{Setup}
\vspace{-0.2cm}
Generally, we follow the settings on image classification tasks in existing passport approaches, including dual-branch DeepIPR~\cite{fan2021deepipr}, Passport-Aware Normalization (PAN)~\cite{zhang2020passport}, and Trapdoor Normalization (TdN)~\cite{liu2023trapdoor}. AlexNet~\cite{krizhevsky2012imagenet} and ResNet-18~\cite{he2016deep}, with Batch Normalization (BN)~\cite{ioffe2015batch} and Group Normalization (GN)~\cite{wu2018group}, are employed to evaluate and compare the performance. The datasets used in the classification experiments include CIFAR-10, CIFAR-100~\cite{krizhevsky2009learning}, Caltech-101, and Caltech-256~\cite{fei2006one}. Besides, DIV2K~\cite{agustsson2017ntire} is used to train the key-based ISN. For passport images $p_\gamma$ and $p_\beta$, we randomly select them from the test set of COCO dataset~\cite{lin2014microsoft}. For steganographic key image $\mathbf{k}_s$, we randomly select it from DTD dataset~\cite{cimpoi2014describing}. For hyper-parameters $\epsilon$, $\omega_s$, and $\omega_b$, we empirically set them as 0.1, 1, and 1. By default, the last three/five normalization layers of AlexNet/ResNet-18 are set as the passport layers.
\begin{table*}[t]
\vspace{-0.0cm}
\centering
\caption{{Inference performance fidelity (in \%) for deployment/verification under ERB ambiguity attacks with BN. FSA and BDR refer to the forged signature accuracy and the bit difference rate between benign and forged signatures, respectively. } }\label{tab:ERB_att}
\vspace{-0.4cm}
\resizebox{\linewidth}{!}{
\begin{tabular}{c  c c c | c c c | c c c | c c c| c c}
\toprule
 & \multicolumn{3}{c|}{CIFAR-10} & \multicolumn{3}{c|}{CIFAR-100} & \multicolumn{3}{c|}{Caltech-101} & \multicolumn{3}{c|}{Caltech-256} &   &  \\
 \hline
 {ResNet-18} & ERB$\downarrow$  & FSA$\downarrow$ &  BDR$\downarrow$ & ERB$\downarrow$  & FSA$\downarrow$ &  BDR$\downarrow$ & ERB$\downarrow$  & FSA$\downarrow$ &  BDR$\downarrow$ & ERB$\downarrow$  & FSA$\downarrow$ &  BDR$\downarrow$ & Mean$\downarrow$ & AD$\uparrow$\\
\hline

DeepIPR & 94.19 / 94.39 & 100 & 50.22 & 62.92 / 64.06  & 100 & 49.49 & 73.56 / 73.89 & 100 & 50.23 & 46.05 / 46.39 & 100 & 49.37 & 69.18 / 69.68 & 0.50 \\

PAN & 94.64 / 94.72 & 100 & 48.71 & 75.28 / 75.10  & 100 & 49.57 & 73.11 / 73.05 & 100 & 49.30 & 53.62 / 53.58 & 100 & 50.08 & 74.16 / 74.11 & 0.90 \\

TdN & 94.48 / 94.24 & 100 & 53.09 & 75.91 / 66.66  & 100 & 48.20 & 73.22 / 64.54 & 99.96 & 50.04 & 54.49 / 34.61 & 100 & 51.76 & 74.53 / 65.01 & 9.51 \\

Ours & 94.70 / \textbf{92.87} & 100 & 49.53 & 75.85 / \textbf{50.27}  & \textbf{99.96} & 49.18 & 73.00 / \textbf{50.06} & \textbf{99.92} & 51.05 & 52.87 / 38.44 & 100 & 49.18 & 73.82 / \textbf{57.91} & \textbf{16.20} \\

\hline
\hline

{AlexNet} & ERB$\downarrow $  & FSA$\downarrow $ &  BDR$\downarrow $ & ERB$\downarrow $  & FSA$\downarrow $ &  BDR$\downarrow $ & ERB$\downarrow $  & FSA$\downarrow $ &  BDR$\downarrow $ & ERB$\downarrow $  & FSA$\downarrow $ &  BDR$\downarrow $ & Mean$\downarrow $ & AD$\uparrow $\\
\hline
DeepIPR & 26.54 / 85.72 & 100 & 47.61 & 46.61 / 51.97  & 100 & 50.11 & 63.56 / 67.83 & 100 & 52.65 & 32.61 / 34.34 & 100 & 51.08 & 42.31 / 59.97 & 17.64 \\

PAN &  91.56 / 82.06 & 100 & 52.43 & 68.36 / 60.29  & 100 & 49.87 & 70.37 / 69.17 & 100 & 53.38 & 44.41 / 40.15 & 100 & 52.56 & 68.68 / 62.92 & 5.76 \\

TdN & 90.16 / 53.16 & 100 & 51.33 & 66.96 / 1.65  & 98.96 & 49.78 & 70.26 / 37.19 & 98.96 & 49.96 & 44.14 / 0.28 & 95.31 & 48.24 & 67.88 / 23.07 & 44.81 \\

Ours & 84.90 / \textbf{16.11} & \textbf{99.57} & 50.24 & 67.38 / \textbf{1.02}  & \textbf{96.75} & 49.72 & 71.26 / \textbf{35.86} & \textbf{98.96} & 50.33 & 45.00 / 4.93 & 100 & 51.17 & 67.14 / \textbf{14.48} & \textbf{52.66} \\

\bottomrule
\end{tabular}
}
\vspace{-0.5cm}
\end{table*}

\subsection{Verification assessment}\label{sec:ver_assess}
We evaluate the effectiveness of the proposed steganographic passport by the verification and inference performances. All the models are trained with 200 epochs, utilizing a multi-step learning rate that decreases gradually from 0.01 to 0.0001. A twofold verification is adopted in our method, which entails a) an ownership verification $\{ \mathcal{V}^\mathcal{I}_{O}, \mathcal{V}^f_{O}, \mathcal{V}^s_{O} \}$ and b) a license verification $\{\mathcal{V}^p_{L},\mathcal{V}^{\mathrm{I}}_{L}\}$. 

For ownership verification, our steganographic passport exhibits the lowest AD values of 0.00\% on ResNet-18 and 0.01\% on AlexNet for $\mathcal{V}^\mathcal{I}_{O}$ with the correct owner passport, as presented in~\Cref{tab:inference}. This is attributed to the effect of the balance loss function, which ensures that the updates of the affine factors of the two branches are more synchronized during training. For $\mathcal{V}^f_{O}$, due to the stronger anti-forgery passport constraint imposed by our method, there is a slight decrease in model inference performance compared to TdN and PAN, but it still outperforms DeepIPR. As shown in~\Cref{tab:inference}, the average deployment and verification branch accuracies of our method on ResNet-18 are both 71.73\%, which is only 0.40\% and 0.07\% lower than the accuracies of the deployment and verification branches, respectively of TdN. The performance of our method peaks in several cases highlighted with the bold-printed percentages. For instance, it achieves the top score of 71.98\% for AlexNet on Caltech-101. For $\mathcal{V}^s_{O}$, the SAs of all four methods achieve 100\% in all the settings.
\begin{figure}[t]
  \centering
\includegraphics[width=0.9\linewidth]{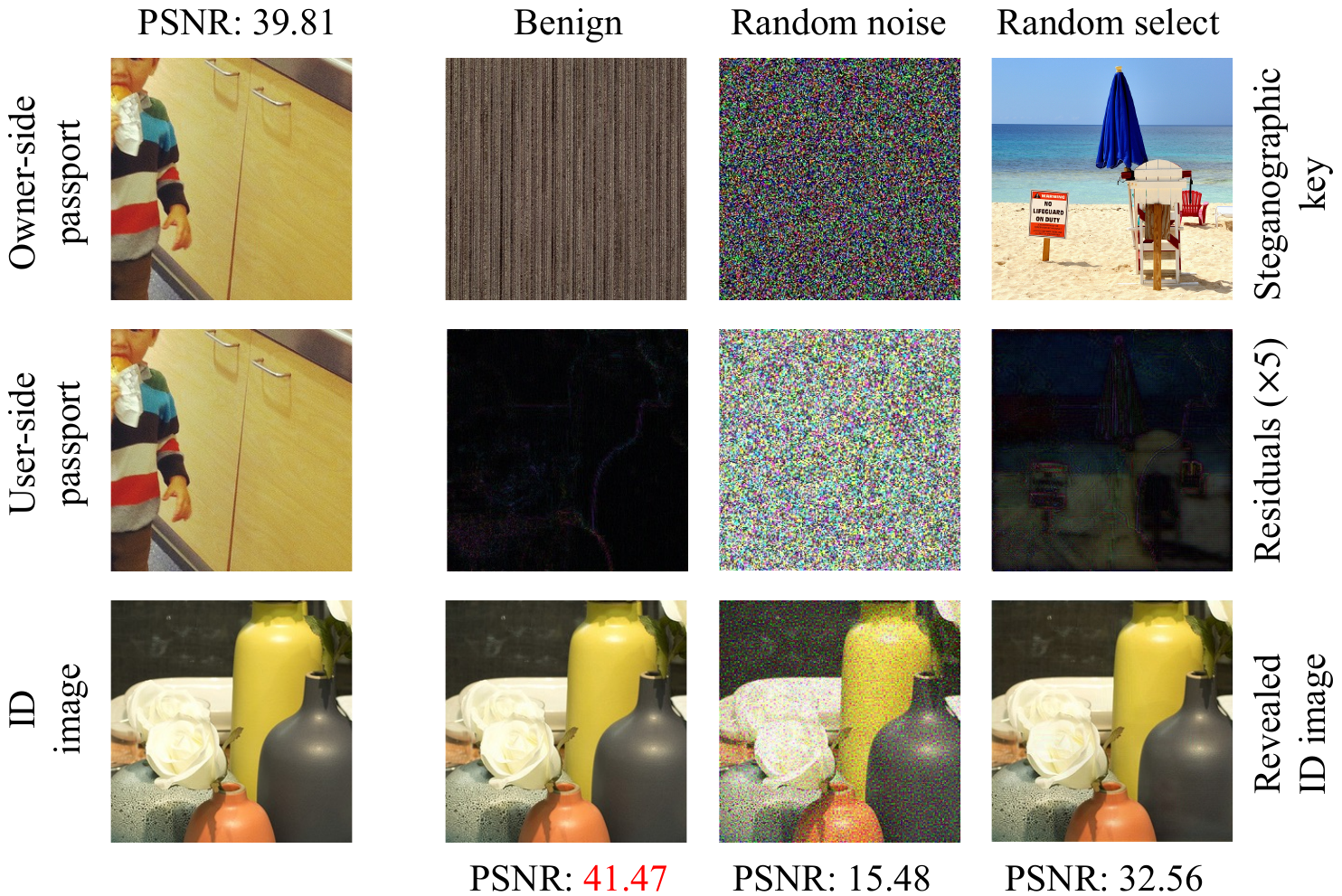}
  \vspace{-0.2cm}
   \caption{{Hiding and revealing performance of the key-based ISN.
   }}
   \label{fig:visual}
   \vspace{-0.6cm}
\end{figure}

For license verification, a comparison of the visual images for $\mathcal{V}^p_{L}$ and  $\mathcal{V}^{\mathrm{I}}_{L}$ is presented in~\cref{fig:visual}. A very high PSNR of 41.47 between the original ID image and the ID image revealed using the genuine steganographic key is obtained, indicating a high level of visual similarity between them. In contrast, using forged keys to reveal the ID image results in significantly lower PSNR. Specifically, a PSNR of only 15.48 is obtained for the revealed ID images by using a random noise-like forged passport image and 32.56 by using a steganographic key image selected randomly from the same distribution as the genuine key image. Furthermore, upon repeating this evaluation 100 times, the average PSNRs obtained from revealing the ID image by using random noise-like forged and randomly-selected key images are 15.68 and 35.89, respectively.

Based on the above observations, we empirically set $\mathbf{\tau}_{d}$, $\mathbf{\tau}_{\xi}$, $\mathbf{\tau}_{p}$ and $\mathbf{\tau}_{r}$ to 0.05\%, 93\%, 39 and 41 for $\mathcal{V}^\mathcal{I}_{O}$, $\mathcal{V}^s_{O}$, $\mathcal{V}^p_{L}$ and $\mathcal{V}^{\mathrm{I}}_{L}$, respectively. For $\mathcal{V}^f_{O}$, $\mathbf{\tau}_{f}$ is calibrated according to the task as the same clean model trained for different tasks has different inference performance.

\vspace{-0.2cm}
\subsection{Robustness against ownership ambiguity attacks}\label{sec:own_amb_att}
\vspace{-0.2cm}
Ownership ambiguity attacks aim to falsely claim the ownership of a deep model by satisfying the conditions outlined in \textit{Remark 1}. Owing to the avalanche criterion provided by the hash function, attacks that attempt to forge the genuine passport images by random selection~\cite{fan2021deepipr, zhang2020passport, liu2023trapdoor} or tamper with the user-side passport images will fail on $\mathcal{V}_{O}^{s}$ due to the significant signature deviation. Additionally, it is computationally intractable to reverse the uniformly random and collision-resistant cryptographically secure hash function to recover genuine passport images~\cite{liu2023trapdoor}. Thus, we specifically focus on the state-of-the-art ERB ownership ambiguity attack, under the strict assumption that the attacker has stolen both the deployment and verification branches, along with 10\% of original training data. The individual version of ERB, \textit{i.e.}, IERB is used to attack the models with BN.

ERB demonstrated its effectiveness by retraining the model to fit a forged passport. In ERB’s original configuration, attackers directly fabricate the passport layer's affine factors, leveraging the network's differentiable architecture to deduce the passport images. However, this possibility is denied by the hash functions of our method and TdN. Therefore, while maintaining the original settings of ERB for DeepIPR and PAN, the passport layer's affine factors in our method and TdN are to be derived from the corresponding hash-related functions. \Cref{tab:ERB_att} displays the experimental results. 
To successfully meet the criteria of \textit{Remark~1}, the attack must achieve 100\% of forged SA (FSA) for $\mathcal{V}^s_O$. As evinced by the results presented in~\cref{sec:ver_assess}, the legitimate owner is able to comply with this stringent criteria. The experimental results show that the attacks on TdN and our method can only achieve 100\% FSA for four and three model-dataset cases, respectively. Conversely, despite having around 50\% BDR between the forged and genuine signatures, DeepIPR and PAN fail to withstand this attack with 100\% FSA on all the model-dataset combinations. Besides, our method consistently produces the lowest mean performance fidelity of the verification branch for $\mathcal{V}^f_O$, especifically for ResNet-18 (57.91\%) and AlexNet (14.48\%). Furthermore, our method also exhibits the highest AD that surpasses $\mathbf{\tau}_{d}$ of $\mathcal{V}^d_O$ by a significant margin. The results, taken holistically, corroborate that our method has the highest resilience against ambiguity attacks.

\subsection{Robustness against license ambiguity attacks }
\vspace{-0.2cm}
Two license ambiguity attacks are considered. The first attack is to reveal the attacker's ID image by manipulating the genuine user-side passports. The second attack is to reveal the attacker's ID image from the genuine user-side passport image by forging the steganographic key image.
\begin{figure}[!t]
    \centering
    \vspace{-0.1cm}
    \begin{subfigure}{0.232\textwidth}\label{fig:auth_amb_att_a}
   \includegraphics[width=\textwidth]{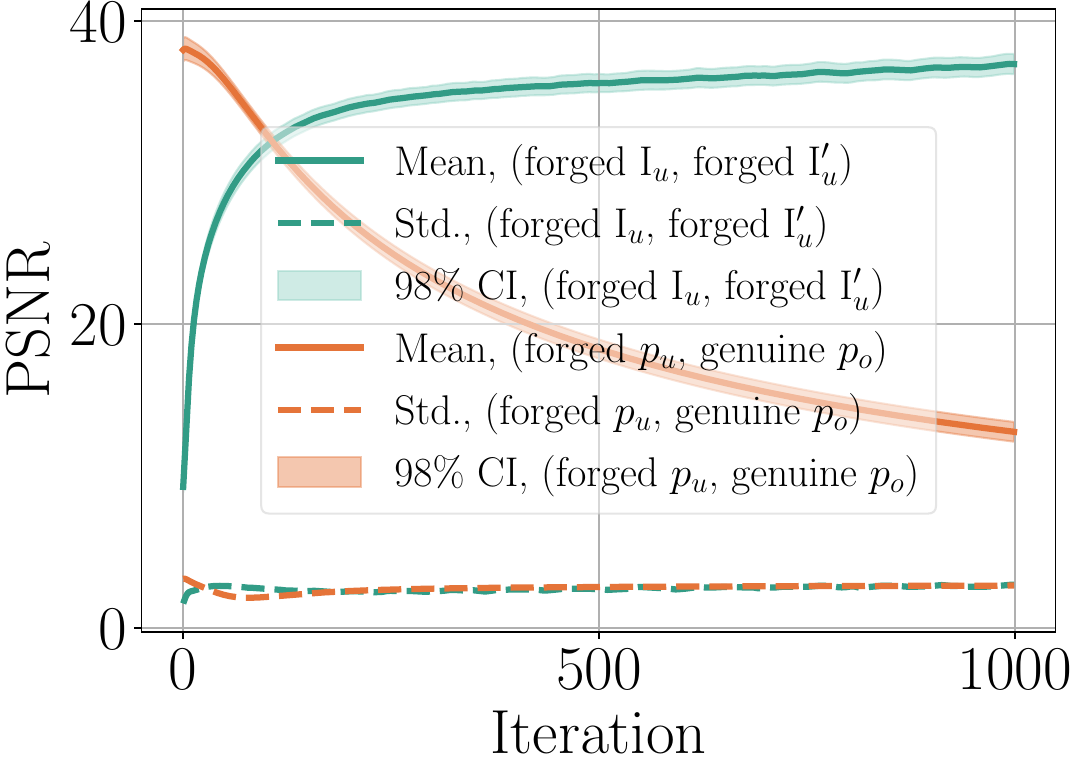}
        \caption{Forged passport}
        \label{fig:auth_amb_att-a}
    \end{subfigure}
    %\hfill 
    \begin{subfigure}{0.232\textwidth} 
\includegraphics[width=\textwidth]{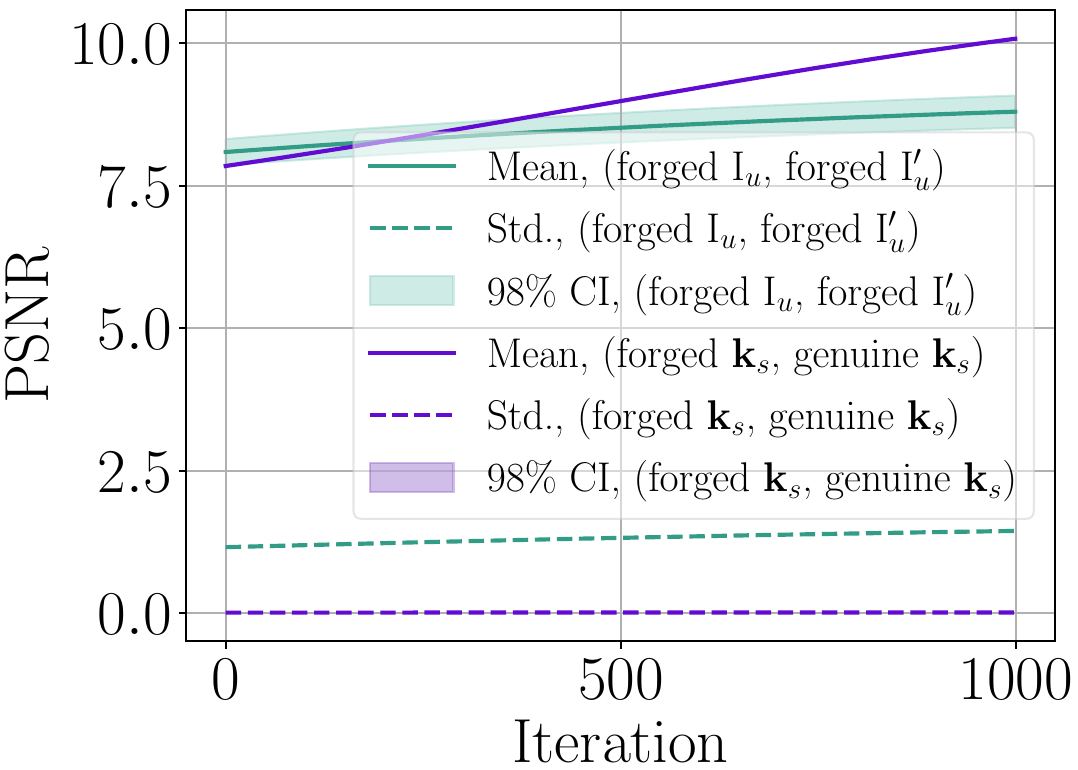}
        \caption{Forged steganographic key}
        \label{fig:auth_amb_att-b}
    \end{subfigure}
    \vspace{-0.4cm}
    \caption{Results of license ambiguity attacks on our method. Batch Normalization is used in the evaluated models. 
    We set the Z-score as 2.33 for 98\% confidence interval (CI).
    }
    \label{fig:auth_amb_att}
    \vspace{-0.6cm}
\end{figure}

\textbf{Forged user-side passport.}
To simulate this attack, we retain the ISN's weights and forge user-side passports with the attacker's ID images by minimizing the reveal loss $\mathcal{L}_r$ between the original forged ID images and revealed forged ID images. The attack was performed by randomly sampling 100 owner-side passport images to generate 100 corresponding user-side passport images, and then using the Adam optimizer on each user-side passport image for 1000 iterations with a learning rate of 0.001. The progression of the attack is depicted in~\cref{fig:auth_amb_att-a}. The mean PSNR between the original forged ID images and their revealed forged ID images increases and the mean PSNR between the forged user-side passport images and genuine owner-side passport images decreases with iterations. As a result, it becomes impossible for the attacker to simultaneously satisfy the criteria of both $\mathcal{V}_{L}^{\mathbf{I}}$ and $\mathcal{V}_{L}^p$.

\textbf{Forged steganographic key.}
In this experiment, the attacker steals the key-based ISN to forge the steganographic key image. We initialize this attack by randomly sampling a noise vector from a Gaussian distribution 100 times. Correspondingly, we also randomly sample a target forged ID image from the distribution of genuine ID images 100 times, ensuring that there is no overlap between the forged and real images. The forged key images are obtained by an Adam optimizer with a learning rate of 0.001 over 1000 iterations. The progress of this process is depicted in~\cref{fig:auth_amb_att-b}, where the shaded region marks the range of average PSNRs obtained over 100 rounds of attack. For the forged key images, the deviation is small and the purple shaded region is not visible for the scale of the plot. These observations indicate that this attack fails to satisfy the criteria of $\mathcal{V}_{L}^{\mathbf{I}}$.

\begin{table*}[t]
\centering
\vspace{-0.8cm}
\caption{{{The performance fidelity (in \%) of deployment model under fine-tuning attacks. The value in bracket is the signature accuracy.}} }\label{tab:finetune}
\vspace{-0.4cm}
\resizebox{\linewidth}{!}{
\begin{tabular}{c  c  c | c  c | c  c | c  c| c }
\toprule
\multirow{2}{*}{ResNet-18} & \multicolumn{2}{c|}{CIFAR-10 to CIFAR-100} & \multicolumn{2}{c|}{CIFAR-100 to CIFAR-10} & \multicolumn{2}{c|}{Caltech-101 to Caltech-256} & \multicolumn{2}{c|}{Caltech-256 to Caltech-101} & \multirow{2}{*}{Mean}   \\
 \cline{2-9}
 & BN  & GN & BN & GN & BN & GN & BN & GN & \\
\hline

DeepIPR & 73.81 (96.16) & 70.71 (95.23) & 92.10 (99.10) & 90.98 (99.45) & 48.53 (99.49) & 41.48 (99.30) & 76.05 (99.92) & 71.75 (100) & 70.68 (98.58) \\

PAN & 71.73 (74.30) & 68.15 (96.37) & 91.97 (78.71) & 90.46 (99.34) & 47.61 (81.29) & 39.51 (84.53) & 79.96 (100) & 71.64 (99.97) & 70.13 (89.31)\\

TdN & 70.77 (56.05)  & 69.43 (62.70)  &  91.86 (87.50)  & 89.93 (73.87) &  46.55 (90.51) & 39.88 (82.30)  & 77.12 (96.09) & 71.58 (82.81) & 69.64 (78.98)\\

Ours & 64.13 (98.60)  & 65.11 (93.91) & 92.28 (96.17)  & 90.22 (93.86) & 43.91 (98.24)  & 37.48 (97.11) & 76.09 (99.09) & 69.04 (99.26) & 
{67.28} (97.03) \\

\bottomrule
\end{tabular}
}
\vspace{-0.5cm}
\end{table*}

\subsection{Robustness against removal attacks }
\vspace{-0.2cm}
We consider removal attacks that happen on the deployment branch when that branch is stolen by the attacker or abused by the licensed user. Unlike the pre-deployment model ownership claim in Sec~\ref{sec:own_amb_att}, the model may be deliberately or maliciously altered after its deployment without compromising its utility, $\tau_\xi$ for $\mathcal{V}^s_O$ needs not be infinitely close to 1 to claim the ownership of the deployed model.
\begin{figure}[!t]
    \centering
    \vspace{-0.0cm}
    \begin{subfigure}{0.236\textwidth} 
        \includegraphics[width=\textwidth]{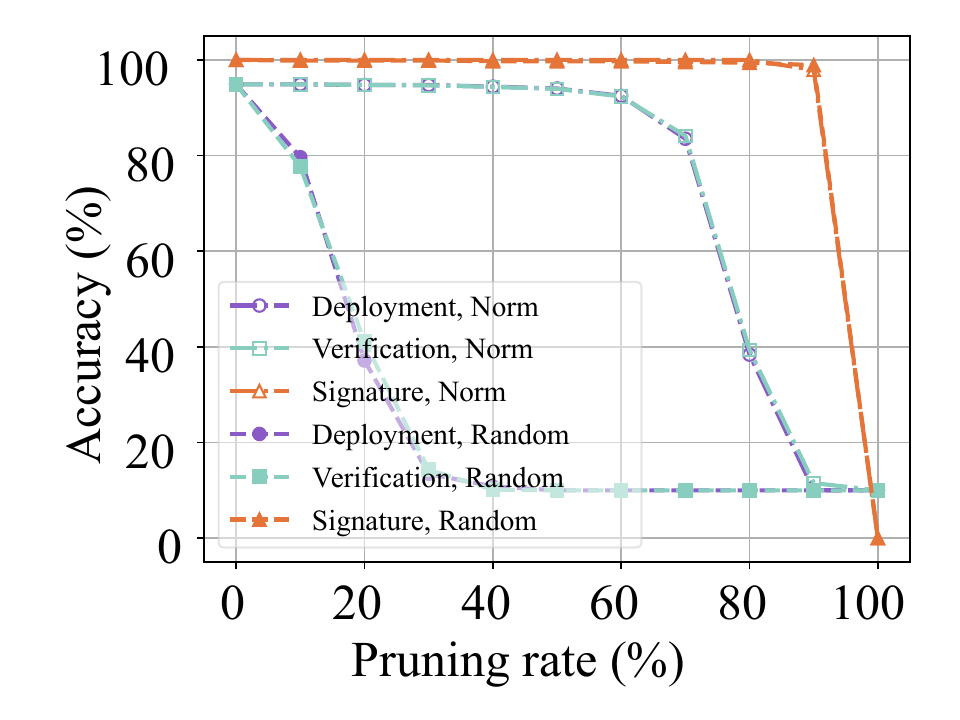}
        \caption{CIFAR-10 with BN}
        \label{fig:prune-a}
    \end{subfigure}
    \begin{subfigure}{0.236\textwidth}
        \includegraphics[width=\textwidth]{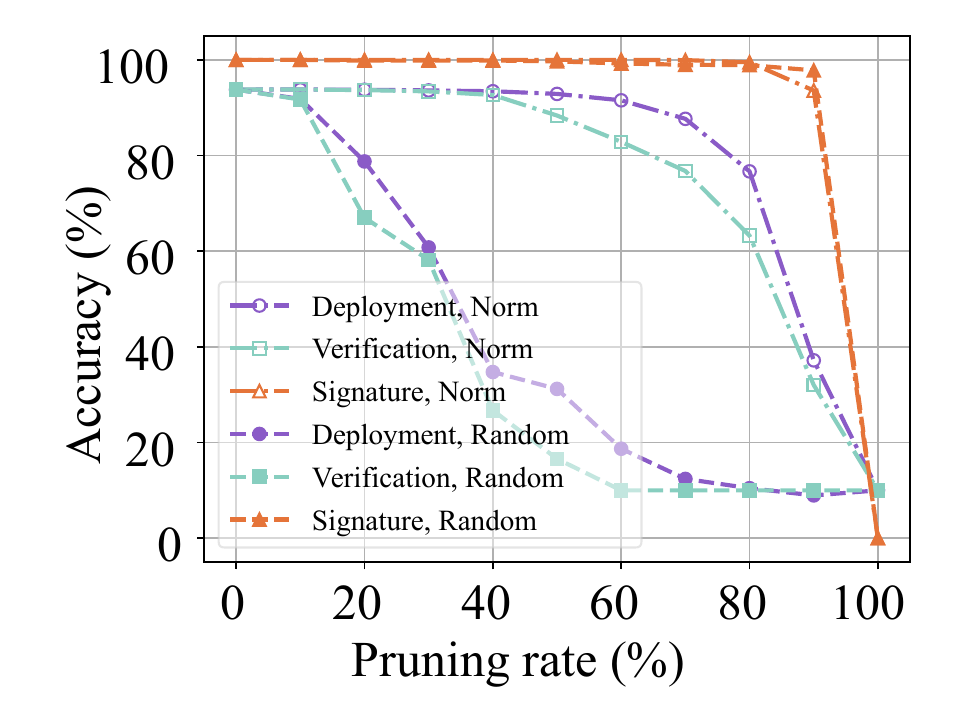}
        \caption{CIFAR-10 with GN}
        \label{fig:prune-b}
    \end{subfigure}
     \vspace{-0.7cm}
    \caption{The performance of our method under random and $\ell_1$ norm pruning attacks.}
    \label{fig:prune}
    \vspace{-0.7cm}
\end{figure}

\textbf{Fine-tuning.}
We consider a transfer-learning task by fine-tuning the deployment branch of a pretrained dual-branch model for 100 epochs at a learning rate of 0.001. As shown in~\Cref{tab:finetune}, our method has the second-highest average SA of 97.03\%. Given the avalanche criterion of the hash function, a false passport will lead to substantially high signature bit errors. The model owner can still attest the ownership under $\mathcal{V}^s_{O}$ with this small signature error. This is because both $\mathcal{V}^f_{O}$ and $\mathcal{V}^\mathcal{I}_{O}$ can still be fulfilled by applying this slightly mismatched signature obtained from the fine-tuned deployment branch to the original dual branch model kept by the model owner. Compared to other passport-based methods, our method has the lowest average performance fidelity upon fine-tuning, making it less attractive to attack our model.

\textbf{Pruning.}
Two prune strategies, \textit{i.e.}, random pruning and $\ell_1$-norm pruning are evaluated. For each model, the pruning rate was increased from 0\% to 100\% with a step size of 10\%. The results are shown in~\cref{fig:prune}. The SA remains relatively high and stable even when the model performance starts to drop abruptly. For example, in~\cref{fig:prune-a}, a 100\% signature detection accuracy is preserved at 80\% pruning rate while the accuracy of the pruned model has fallen to around 50\%. This implies that the pruned model can still pass $\mathcal{V}^f_{O}$ and $\mathcal{V}^s_{O}$ unless the pruning rate is so high that it renders the model unusable. Additionally, the large AD between the two branches also signifies that the model has been modified, as stated in~\cref{sec:ownership_veri}.
 
\subsection{Ablation study}
\vspace{-0.2cm}
This section investigates the variations of our method without the TLP structure and with other activation functions.

\textbf{Without TLP.}
We examine the inference performance fidelity when the TLP is removed from the BN variants of ResNet-18 and AlexNet, and compare their performances across different datasets. The results presented in~\Cref{tab:ablation_TLP} indicate that the drop in inference performance is trivial for models without TLP compared to their corresponding  TLP counterparts. The higher AD of AlexNet, 0.23\% as opposed to 0.00\% of ResNet-18, implies that TLP has negligible impact on the performance fidelity of our method for complex networks but is essential to keep for simpler networks.

\begin{table}[t]
\centering
\caption{{ Inference performance fidelity (in \%) for the deployment/verification branches of our method without TLP.} }\label{tab:ablation_TLP}
\vspace{-0.4cm}
\resizebox{\linewidth}{!}{
\begin{tabular}{c  c   | c   | c   | c  | c }
\toprule
 & {CIFAR-10} & {CIFAR-100} & {Caltech-101} & {Caltech-256} &    \\
 \cline{1-5}
 \multirow{2}{*}{ResNet-18} & BN$\uparrow$  &  BN$\uparrow$ & BN$\uparrow$ & BN$\uparrow$ & AD$\downarrow$\\
\cline{2-6}
  & 94.47 / 94.47  & 75.19 / 75.20 & 71.53 / 71.53 & 50.46 / 50.46 & 0.00   \\
\hline
\hline
\multirow{2}{*}{AlexNet}& BN$\uparrow$  &  BN$\uparrow$ & BN$\uparrow$ & BN$\uparrow$ & AD $\downarrow$\\
  \cline{2-6}
  & 91.05 / 91.25  & 68.26 / 68.65 & 70.11 / 70.11 & 42.42 / 42.75 & 0.23  \\
\bottomrule
\end{tabular}
}
\vspace{-0.3cm}
\end{table}
\begin{table}[t]
    \centering
    \caption{{Performance (in \%) integrating Sigmoid and LeakyReLU in our proposed passport architecture, trained on CIFAR-10.
    Dep/Ver refer to deployment/verification branch, respectively.
} }\label{tab:ablation_activation}
    \vspace{-0.4cm}
    \resizebox{\linewidth}{!}{
    \begin{tabular}{c  c c c | c c c }
    \toprule
     \multirow{2}{*}{ResNet-18-BN} & \multicolumn{3}{c|}{Inference} & \multicolumn{3}{c}{Ambiguity attack (ERB)} \\
     \cline{2-7}
      & Dep / Ver  & SA & AD & Dep / Ver & FSA & AD \\
    \hline
    
    Sigmoid & 94.60 / 94.66 & 100 & 0.06 & 94.49 / 17.76 & 99.84 & 76.73   \\
    
    LeakyReLU & 94.96 / 94.55 & 100 & 0.41 & 94.46 / 91.93 & 100 & 2.53  \\
    
    \bottomrule
    \end{tabular}
    }
    \vspace{-0.6cm}
\end{table}

\textbf{With other activation functions.}
To assess the impact of different activation functions on our method, we also incorporate Sigmoid and LeakyReLU into ResNet-18 with BN. For each activation function, the inference performance and robustness to the ERB attack are evaluated and presented in~\Cref{tab:ablation_activation}. Both activation functions demonstrate comparable performance fidelity between the deployment and verification branches, with SA of 100\% for both activation functions, and AD of 0.06\% and 0.41\% for Sigmoid and LeakyReLU, respectively. The Sigmoid-based model is more resilient to ERB attack than the LeakyReLU-based model, with a significantly higher AD of 76.73\%. Nonetheless, we can adapt the tamper threshold $\tau_d$ to the activation function for higher ambiguity attack resistance.

\vspace{-0.2cm}
\section{Conclusion}
\vspace{-0.2cm}
In this paper, we introduce Steganographic Passport, a novel passport-based model IP licensing protection method that allows the original model ownership to be verified with the passport-aware branch and individual licensees to be verified from the passport-free deployed model without requiring retraining the model for the admission of each licensed user. Legal license verification is made possible by hiding and revealing the licensed user’s ID image in a user-side passport of the deployed model by the invertible steganographic network without compromising the accuracy and security of original model ownership verification. Activation-level obfuscation is added to heighten the security of both passports against ambiguity attacks. The affine factors of the deployment and verification branches, as well as the signatures and model’s weights are tightly coupled during training to safeguard the owner-side passport and the deployed model’s weights against malicious tampering to succeed in forging the ownership or licenseship. The experimental results substantiate the resilience of our method against a range of attacks, including ownership ambiguity attacks, license ambiguity attacks, and removal attacks.

{
    \small
    \bibliographystyle{ieeenat_fullname}
    \bibliography{main}

\begin{thebibliography}{31}
\providecommand{\natexlab}[1]{#1}
\providecommand{\url}[1]{\texttt{#1}}
\expandafter\ifx\csname urlstyle\endcsname\relax
  \providecommand{\doi}[1]{doi: #1}\else
  \providecommand{\doi}{doi: \begingroup \urlstyle{rm}\Url}\fi

\bibitem[Agustsson and Timofte(2017)]{agustsson2017ntire}
Eirikur Agustsson and Radu Timofte.
\newblock Ntire 2017 challenge on single image super-resolution: Dataset and
  study.
\newblock In \emph{Proceedings of the IEEE conference on computer vision and
  pattern recognition workshops}, pages 126--135, 2017.

\bibitem[Chen et~al.(2020)Chen, Dai, Liu, Chen, Yuan, and Liu]{chen2020dynamic}
Yinpeng Chen, Xiyang Dai, Mengchen Liu, Dongdong Chen, Lu Yuan, and Zicheng
  Liu.
\newblock Dynamic {ReLU}.
\newblock In \emph{European Conference on Computer Vision}, pages 351--367.
  Springer, 2020.

\bibitem[Chen et~al.(2023)Chen, Tian, Chen, and Zhou]{chen2023effective}
Yiming Chen, Jinyu Tian, Xiangyu Chen, and Jiantao Zhou.
\newblock Effective ambiguity attack against passport-based dnn intellectual
  property protection schemes through fully connected layer substitution.
\newblock In \emph{Proceedings of the IEEE/CVF Conference on Computer Vision
  and Pattern Recognition}, pages 8123--8132, 2023.

\bibitem[Cimpoi et~al.(2014)Cimpoi, Maji, Kokkinos, Mohamed, and
  Vedaldi]{cimpoi2014describing}
Mircea Cimpoi, Subhransu Maji, Iasonas Kokkinos, Sammy Mohamed, and Andrea
  Vedaldi.
\newblock Describing textures in the wild.
\newblock In \emph{Proceedings of the IEEE conference on computer vision and
  pattern recognition}, pages 3606--3613, 2014.

\bibitem[Fan et~al.(2019)Fan, Ng, and Chan]{fan2019rethinking}
Lixin Fan, Kam~Woh Ng, and Chee~Seng Chan.
\newblock Rethinking deep neural network ownership verification: Embedding
  passports to defeat ambiguity attacks.
\newblock \emph{Advances in neural information processing systems}, 32, 2019.

\bibitem[Fan et~al.(2021)Fan, Ng, Chan, and Yang]{fan2021deepipr}
Lixin Fan, Kam~Woh Ng, Chee~Seng Chan, and Qiang Yang.
\newblock {DeepIPR}: Deep neural network ownership verification with passports.
\newblock \emph{IEEE Transactions on Pattern Analysis and Machine
  Intelligence}, 44\penalty0 (10):\penalty0 6122--6139, 2021.

\bibitem[Feng and Zhang(2020)]{feng2020watermarking}
Le Feng and Xinpeng Zhang.
\newblock Watermarking neural network with compensation mechanism.
\newblock In \emph{Knowledge Science, Engineering and Management: 13th
  International Conference, KSEM 2020, Hangzhou, China, August 28--30, 2020,
  Proceedings, Part II 13}, pages 363--375. Springer, 2020.

\bibitem[Guan et~al.(2020)Guan, Feng, Zhang, Zhou, Zhang, and
  Yu]{guan2020reversible}
Xiquan Guan, Huamin Feng, Weiming Zhang, Hang Zhou, Jie Zhang, and Nenghai Yu.
\newblock Reversible watermarking in deep convolutional neural networks for
  integrity authentication.
\newblock In \emph{Proceedings of the 28th ACM International Conference on
  Multimedia}, pages 2273--2280, 2020.

\bibitem[He et~al.(2016)He, Zhang, Ren, and Sun]{he2016deep}
Kaiming He, Xiangyu Zhang, Shaoqing Ren, and Jian Sun.
\newblock Deep residual learning for image recognition.
\newblock In \emph{Proceedings of the IEEE conference on computer vision and
  pattern recognition}, pages 770--778, 2016.

\bibitem[He et~al.(2019)He, Zhang, and Lee]{he2019sensitive}
Zecheng He, Tianwei Zhang, and Ruby Lee.
\newblock Sensitive-sample fingerprinting of deep neural networks.
\newblock In \emph{Proceedings of the IEEE/CVF conference on computer vision
  and pattern recognition}, pages 4729--4737, 2019.

\bibitem[Hua et~al.(2023)Hua, Teoh, Xiang, and Jiang]{hua2023unambiguous}
Guang Hua, Andrew Beng~Jin Teoh, Yong Xiang, and Hao Jiang.
\newblock Unambiguous and high-fidelity backdoor watermarking for deep neural
  networks.
\newblock \emph{IEEE Transactions on Neural Networks and Learning Systems},
  2023.

\bibitem[Ioffe and Szegedy(2015)]{ioffe2015batch}
Sergey Ioffe and Christian Szegedy.
\newblock Batch {N}ormalization: Accelerating deep network training by reducing
  internal covariate shift.
\newblock In \emph{International conference on machine learning}, pages
  448--456. pmlr, 2015.

\bibitem[Javadi et~al.(2021)Javadi, Norval, Cloete, and
  Singh]{javadi2021monitoring}
Seyyed~Ahmad Javadi, Chris Norval, Richard Cloete, and Jatinder Singh.
\newblock Monitoring {AI} services for misuse.
\newblock In \emph{Proceedings of the 2021 AAAI/ACM Conference on AI, Ethics,
  and Society}, pages 597--607, 2021.

\bibitem[Jia et~al.(2021)Jia, Choquette-Choo, Chandrasekaran, and
  Papernot]{jia2021entangled}
Hengrui Jia, Christopher~A Choquette-Choo, Varun Chandrasekaran, and Nicolas
  Papernot.
\newblock Entangled watermarks as a defense against model extraction.
\newblock In \emph{30th USENIX Security Symposium (USENIX Security 21)}, pages
  1937--1954, 2021.

\bibitem[Jing et~al.(2021)Jing, Deng, Xu, Wang, and Guan]{jing2021hinet}
Junpeng Jing, Xin Deng, Mai Xu, Jianyi Wang, and Zhenyu Guan.
\newblock {HiNet}: deep image hiding by invertible network.
\newblock In \emph{Proceedings of the IEEE/CVF international conference on
  computer vision}, pages 4733--4742, 2021.

\bibitem[Krizhevsky et~al.(2009)Krizhevsky, Hinton,
  et~al.]{krizhevsky2009learning}
Alex Krizhevsky, Geoffrey Hinton, et~al.
\newblock Learning multiple layers of features from tiny images.
\newblock 2009.

\bibitem[Krizhevsky et~al.(2012)Krizhevsky, Sutskever, and
  Hinton]{krizhevsky2012imagenet}
Alex Krizhevsky, Ilya Sutskever, and Geoffrey~E Hinton.
\newblock Imagenet classification with deep convolutional neural networks.
\newblock \emph{Advances in neural information processing systems}, 25, 2012.

\bibitem[Li et~al.(2006)Li, Fergus, and Perona]{fei2006one}
Fei-Fei Li, Robert Fergus, and Pietro Perona.
\newblock One-shot learning of object categories.
\newblock \emph{IEEE transactions on pattern analysis and machine
  intelligence}, 28\penalty0 (4):\penalty0 594--611, 2006.

\bibitem[Lin et~al.(2014)Lin, Maire, Belongie, Hays, Perona, Ramanan,
  Doll{\'a}r, and Zitnick]{lin2014microsoft}
Tsung-Yi Lin, Michael Maire, Serge Belongie, James Hays, Pietro Perona, Deva
  Ramanan, Piotr Doll{\'a}r, and C~Lawrence Zitnick.
\newblock Microsoft {COCO}: Common objects in context.
\newblock In \emph{Computer Vision--ECCV 2014: 13th European Conference,
  Zurich, Switzerland, September 6-12, 2014, Proceedings, Part V 13}, pages
  740--755. Springer, 2014.

\bibitem[Liu et~al.(2023)Liu, Weng, Zhu, and Mu]{liu2023trapdoor}
Hanwen Liu, Zhenyu Weng, Yuesheng Zhu, and Yadong Mu.
\newblock Trapdoor normalization with irreversible ownership verification.
\newblock In \emph{International Conference on Machine Learning}, pages
  22177--22187. PMLR, 2023.

\bibitem[Liu et~al.(2024)Liu, Ye, Chen, and Lam]{liu2024threats}
Ziyao Liu, Huanyi Ye, Chen Chen, and Kwok-Yan Lam.
\newblock Threats, attacks, and defenses in machine unlearning: A survey.
\newblock \emph{arXiv preprint arXiv:2403.13682}, 2024.

\bibitem[Lu et~al.(2021)Lu, Wang, Zhong, and Rosin]{lu2021large}
Shao-Ping Lu, Rong Wang, Tao Zhong, and Paul~L Rosin.
\newblock Large-capacity image steganography based on invertible neural
  networks.
\newblock In \emph{Proceedings of the IEEE/CVF conference on computer vision
  and pattern recognition}, pages 10816--10825, 2021.

\bibitem[Ong et~al.(2021)Ong, Chan, Ng, Fan, and Yang]{ong2021protecting}
Ding~Sheng Ong, Chee~Seng Chan, Kam~Woh Ng, Lixin Fan, and Qiang Yang.
\newblock Protecting intellectual property of generative adversarial networks
  from ambiguity attacks.
\newblock In \emph{Proceedings of the IEEE/CVF Conference on Computer Vision
  and Pattern Recognition}, pages 3630--3639, 2021.

\bibitem[Peng et~al.(2022)Peng, Li, Chen, Zhang, Zhu, and
  Xue]{peng2022fingerprinting}
Zirui Peng, Shaofeng Li, Guoxing Chen, Cheng Zhang, Haojin Zhu, and Minhui Xue.
\newblock Fingerprinting deep neural networks globally via universal
  adversarial perturbations.
\newblock In \emph{Proceedings of the IEEE/CVF Conference on Computer Vision
  and Pattern Recognition}, pages 13430--13439, 2022.

\bibitem[Quan et~al.(2023)Quan, Teng, Xu, Huang, and
  Ji]{quan2023fingerprinting}
Yuhui Quan, Huan Teng, Ruotao Xu, Jun Huang, and Hui Ji.
\newblock Fingerprinting deep image restoration models.
\newblock In \emph{Proceedings of the IEEE/CVF International Conference on
  Computer Vision}, pages 13285--13295, 2023.

\bibitem[Sanadhya and Sarkar(2008)]{sanadhya2008new}
Somitra~Kumar Sanadhya and Palash Sarkar.
\newblock New collision attacks against up to 24-step sha-2.
\newblock In \emph{Progress in Cryptology-INDOCRYPT 2008: 9th International
  Conference on Cryptology in India, Kharagpur, India, December 14-17, 2008.
  Proceedings 9}, pages 91--103. Springer, 2008.

\bibitem[Shafieinejad et~al.(2021)Shafieinejad, Lukas, Wang, Li, and
  Kerschbaum]{shafieinejad2021robustness}
Masoumeh Shafieinejad, Nils Lukas, Jiaqi Wang, Xinda Li, and Florian
  Kerschbaum.
\newblock On the robustness of backdoor-based watermarking in deep neural
  networks.
\newblock In \emph{Proceedings of the 2021 ACM Workshop on Information Hiding
  and Multimedia Security}, pages 177--188, 2021.

\bibitem[Uchida et~al.(2017)Uchida, Nagai, Sakazawa, and
  Satoh]{uchida2017embedding}
Yusuke Uchida, Yuki Nagai, Shigeyuki Sakazawa, and Shinichi Satoh.
\newblock Embedding watermarks into deep neural networks.
\newblock In \emph{Proceedings of the 2017 ACM on international conference on
  multimedia retrieval}, pages 269--277, 2017.

\bibitem[Wang et~al.(2022)Wang, Xu, Zheng, and Chang]{wang2022buyer}
Si Wang, Chaohui Xu, Yue Zheng, and Chip-Hong Chang.
\newblock A buyer-traceable dnn model {IP} protection method against piracy and
  misappropriation.
\newblock In \emph{2022 IEEE 4th International Conference on Artificial
  Intelligence Circuits and Systems (AICAS)}, pages 308--311. IEEE, 2022.

\bibitem[Wu and He(2018)]{wu2018group}
Yuxin Wu and Kaiming He.
\newblock Group normalization.
\newblock In \emph{Proceedings of the European conference on computer vision
  (ECCV)}, pages 3--19, 2018.

\bibitem[Zhang et~al.(2020)Zhang, Chen, Liao, Zhang, Hua, and
  Yu]{zhang2020passport}
Jie Zhang, Dongdong Chen, Jing Liao, Weiming Zhang, Gang Hua, and Nenghai Yu.
\newblock Passport-aware normalization for deep model protection.
\newblock \emph{Advances in Neural Information Processing Systems},
  33:\penalty0 22619--22628, 2020.

\end{thebibliography}
}


\end{document}